\documentclass[preprint,9pt]{elsarticle}

\usepackage{amssymb}

\usepackage[nodots,nocompress]{numcompress}

\biboptions{square}

\journal{Nuclear Physics B}

\begin{document}

\begin{frontmatter}



\title{Gauge Symmetries and Holographic Anomalies of Chern-Simons and Transgression AdS Gravity}


\author{Pablo Mora}
\ead{pablomora@cure.edu.uy}
\address{Centro Universitario Regional Este (CURE), Universidad de la Rep\'ublica, Uruguay, Ruta 9 Km 207, Rocha, Uruguay}

\begin{abstract}
We review the issue of gauge and gravitational anomalies with backgrounds, 
maybe offering a new outlook on some aspects of these questions.

We compute the holographic anomalies of hypothetical theories dual, in the sense of the AdS-CFT correspondence, to Chern-Simons AdS gravities. Those anomalies are either gauge anomalies associated to the AdS gauge group of the theory or diffeomorphism anomalies, with each kind related to the other.
 
As a result of using suitable action principles por Chern-Simons AdS gravities, coming from Transgression forms, we obtain finite results without the need for further regularization. 

Our results are of potential interest for Lovelock gravity theories, as it has been shown that the boundary terms dictated by the transgressions for Chern-Simons gravities are also suitable to regularize Lovelock theories.  The Wess-Zumino consistency condition ensures that anomalies of the generic form computed here should appear for these and other theories. 
\end{abstract}


\end{frontmatter}















\section{Introduction}

Chern-Simons (CS) gravities have been investigated by many researchers during the last few decades, uncovering a wealth of interesting properties in many aspects of this theories 
(see for instance \cite{chern-simons-2+1-1,chern-simons-2+1-2,chamseddine-1,chamseddine-2,banados-troncoso-zanelli-1,banados-troncoso-zanelli-2}). 
A recent review with a quite complete list of references is \cite{zanelli-lectures}.  

Within the area of research of Chern-Simons gauge and gravity theories, a number of papers was devoted to  passing
from Chern-Simons to Transgression forms as a way to address several issues ith Chern-Simons theories (see for instance
refs.\cite{potsdam,transgression-branes-1,transgression-branes-2,more-transgressions-1,more-transgressions-2,IRS-1,IRS-2,sarda-1,sarda-2,tesis,motz1,motz3}). The work discussed here follows the general strategy 
and outlook advanced in these papers.

The subject of anomalies in quantum field theories is intimately  related, for deep mathematical and physical reasons,
to the study of Chern-Simons forms \cite{wu-zee-zumino,bardeen-zumino, manes-stora-zumino,alvarez}.
We are interested in the possible anomalies of hypothetical theories dual in the sense of the AdS-CFT conjecture
\cite{maldacena,witten-ads,Gubser} to Chern-Simons AdS gravities.
 
Holographic anomalies provided some of the firts tests of the AdS-CFT correspondence \cite{witten-ads,henningson},
verifying the agreement of the Conformal (or Weyl) anomaly \cite{duff-weyl-anomaly} for both sides of the duality.
This check referred to General Relativity with cosmological constant regulated by suitable boundary terms in 5D in the bulk  side and super Yang-Mill theory with ${\cal N}=4$ and a large number of colors (related to the cosmological constant of the bulk) in 4D in the boundary.

After that conformal anomalies induced by higher curvature gravitational theories in 5D where computed following similar methods\cite{nojiri-odintsov,blau-narain-gava,schwimmer-theisen}.

The conformal anomaly induced by Chern-Simons gravitational theories was computed in \cite{banados-schwimmer-theisen} for 5D and 3D CS gravity (inducing Weyl anomalies in 4D and 2D CFTs respectively) and the generic form for arbitrary dimension was conjectured. This calculation was done only on the gravitational side, as the dual CFT theories are not known, but the result of doing this gravitational computation using several different methods was the same, and it was in agreement with what was to be expected from refs.\cite{nojiri-odintsov,blau-narain-gava,schwimmer-theisen}. 
The Weyl anomaly in 4D induced by CS gravity in 5D was also computed in \cite{banados-miskovic-theisen} while the Weyl anomaly induced by CS gravities in any dimension was computed in \cite{banados-olea-theisen}. In those works the Weyl anomaly was computed as the trace of the boundary energy-momentum tensor, adding counterterms to cancel infinite contributions. 
An analysis of Weyl anomalies for generic gravity theories, including CS gravities, was done in ref.\cite{schwimmer-theisen-2}, where it was emphazised in particular that the bulk classical field equations are not required to compute the Weyl anomaly.

In the present work we compute holographic anomalies associated to the AdS gauge invariance of Chern-Simons AdS gravity.
We start from the action principles discussed in \cite{mora-action-principle-cs-ads}, and assume the asymptotic behavior of the
fields assumed in that work\footnote{Analogous in principle to the Fefferman-Graham expansion \cite{fefferman-graham},
but adapted to a first order formulation allowing a non vanishing bulk torsion.}. One important point is that those action principles involve two sets of gauge fields, one of which is regarded as a regulator. The anomaly is then seen as a result of 
not varying the regulator, as the whole transgression action with both sets of fields varying is in fact gauge invariant. 
A consequence of this built in regulation of the action principles considered is that every magnitude of interest is finite from the beginning, without needing further subtractions or corrections\footnote{The present work may be regarded as an improvement and extension of ref.\cite{mora-2010}, which dealt with the Weyl anomaly for CS-AdS gravity, following the same general outlook.}.

The plan of this work is the following:

In Section 2 I review gauge and diffeomorphism anomalies with backgrounds, possibly giving some new point
of view in some aspects of this question.

In Section 3 I give a very brief review of Transgression and AdS gravity.

In Section 4 I analyze the subset of AdS gauge transformations consistent with the 
asymptotic fall-off of the fields discussed in \cite{mora-action-principle-cs-ads}.

In Section 5. AdS holographic gauge anomalies are computed
for one of tha action principles
considered in \cite{mora-action-principle-cs-ads} ("Backgrounds").

In Section 6. we study AdS holographic gauge anomalies for the second action principle
discussed in \cite{mora-action-principle-cs-ads} ("Kounterterms").

In Section 7. and 8. we discuss diffeomorphism anomalies 
for the Backgrounds and Kounterterms action principles respectively.

Section 9. contains discussion of the results presented and conclusions.

\section{Chern-Simons and Transgression forms and Anomalies}

\subsection{Goal of this section}

In this section we review the main results on gauge and gravitational anomalies
with backgrounds, with some mathematical preliminaries,  required in the following sections.
Gauge anomalies with backgrounds were studied by Ma\~nes et al. in \cite{manes-stora-zumino}. 
What 
follows could in principle be 
obtained from that work, nevertheless we find it convenient to derive the 
explicit forms of the anomalies presented
here, that are used below. Some of the those results were presented 
in \cite{mora-2010}. I am also aware of expressions similar to some of the ones 
presented here in the work by Moss \cite{moss}.
Gravitational and Weyl anomalies have been 
discussed in \cite{bardeen-zumino,alvarez} and \cite{duff-weyl-anomaly} respectively,
for instance, but I am not aware of any work presenting diffeomorphism anomalies with backgrounds 
in the way done in subsection 2.4 below.

The content of this section on gauge and gravitational anomalies applies in principle
to any theory involving gauge fields and gravitation, while in the following sections we 
will apply those results to a specific kind of theories (CS AdS gravity).

\subsection{Transgressions}

Chern-Simons forms\footnote{For the details of the mathematics of Chern-Simons and Transgression forms and references see \cite{nakahara}.} $\mathcal{C}_{2n+1}(A)$ are differential forms defined for
a connection $A$, which under gauge transformations of that connection transform by a closed form, so are say to be \textit{quasi invariant}. Transgression forms $\mathcal{T}_{2n+1}$ are a generalization of Chern-Simons forms that depend on two gauge connections $A$ and $\overline{A}$ and are strictly gauge invariant if both connections are subjected to the same gauge transformation.  The use of this forms as lagrangians for physical theories, or as a template for actions for physical theories was discussed in references \cite{motz1, motz3}. Transgressions can be written (see e.g., \cite{nakahara}) as the difference of two Chern-Simons forms plus an exact form 
\begin{equation}
\mathcal{T}_{2n+1}(A,\overline{A})=\mathcal{C}_{2n+1}(A)-\mathcal{C}_{2n+1}(\overline
{A})-d\mathcal{B}_{2n}\left(  A,\overline{A}\right)\label{transgression-1}
\end{equation}
where $\mathcal{T}_{2n+1}(A,\overline{A}=0)=\mathcal{C}_{2n+1}(A)$,  or explicitly as 
\begin{equation}\label{transgression}
\mathcal{T}_{2n+1}\left(  A,\overline{A}\right)  =(n+1)\int_{0}^{1}
dt\ <\Delta{A}F_{t}^{n}>\label{transgression-2}
\end{equation}
where\footnote{Here wedge product between forms is assumed.} 
$A_{t} = tA+(1-t)\overline{A}=\overline{A}+t\Delta A $
is a connection that interpolates between the two independent gauge potentials
$A$ and $\overline{A}$. The Lie algebra-valued one-forms\footnote{Notation: In what follows upper case Latin indices from the beginning of the alphabet $A,~B,~C,...$ are space-time indices with values from $0$ to $d-1=2n$; upper case Latin indices from the middle of the alphabet $I,~J,~K,...$ are space-time indices with values from $0$ to $d-1=2n$ but different from 1 (with 1 corresponding to a "radial" coordinate, or a coordinate along the direction normal to the boundary); lower case Latin indices from the beginning of the alphabet $a,~b,~c,...$ are tangent space (or Lorentz) indices with values from $0$ to $d-1=2n$; lower case Latin indices from the middle of the alphabet $i,~j,~k,...$ are tangent space (or Lorentz) indices with values from $0$ to $d-1=2n$ but different from 1 (with 1 identified to a "radial" direction, or a direction normal to the boundary in tangent space). The index $\alpha$ labels the generators $G_{\alpha}$ of the Lie group considered and takes values from 1 to the dimension of the group.} $A=A_{A}^{\alpha}G_{\alpha}\ dx^{A}$ and $\overline{A}=\overline{A}_{A}^{\alpha}G_{\alpha}\ dx^{A}$ are
connections under gauge transformations, $G_{\alpha}$ are the generators of the gauge group $\mathcal{G}$ (elements of its Lie algebra $\mathfrak{G}$) and
$<\cdots>$ stands for a symmetrized invariant trace in the Lie algebra (or equivalently for the contraction with a symmetric invariant tensor of the group). 
The corresponding curvature is
$F_{t}=dA_{t}+A_{t}^{2}=t F +(1-t)\overline{F}-t(1-t)(\Delta A)^{2}$.  Setting $\overline{A} =0$ 
in the transgression form yields the Chern-Simons form for $A$.
If $g$ is an element of $\mathcal{G}$, then a gauge transformation of $A$ is given by $A^g=g^{-1}[A+d]g$ and the field strength transforms covariantly as $F^g=g^{-1}Fg$. If $\overline{A}$ is transformed with the same group element, then $\Delta A$ and $F_t$ transform covariantly, and from eq.[~\ref{transgression}] it is clear that the transgression is gauge invariant in that case. The case where $A$ is transformed but $\overline{A}$ is not is considered in the next subsection, and it is relevant to compute gauge anomalies with backgrounds.

\subsection{Gauge Anomaly with Background}

The variation of the transgression under infinitesimal variations of $A$ and $\overline{A}$ is
\begin{eqnarray}
\delta\mathcal{T}_{2n+1}=(n+1) <F^n\delta A>-(n+1)<\overline{F}^n\delta \overline{A}>\nonumber\\
-n(n+1)d\{\int _0^1dt<\Delta AF_t^{n-1}\delta A_t>\}\label{variation-transgression}
\end{eqnarray} 
We are interested in how the transgression transforms under infinitesimal gauge transformations $g=1+\lambda$ (with $\lambda$ infinitesimal) that change $A$ but not $\overline{A}$, given by $\delta A=D\lambda$ and $\delta \overline{A}=0$. 
However, to get the variation of the transgression if only $A$ varies it is 
actually easier to exploit the fact that the transgression is invariant if both gauge potentials are varied by the with the same gauge transformation, an then isolating the part that corresponds to only varying $A$. We start then taking $\delta _{\lambda} A=D\lambda =d\lambda +[A,\lambda]$ and $\delta _{\lambda}\overline{A}=\overline{D}\lambda=d\lambda +[\overline{A},\lambda]$
\begin{eqnarray}
0=\delta _{\lambda}\mathcal{T}_{2n+1}=+(n+1) <F^nD\lambda> -(n+1) <\overline{F}^n\overline{D}\lambda> -\nonumber\\
-n(n+1)d\{\int _0^1dt<\Delta AF_t^{n-1}\delta _{\lambda}A_t>\} 
\end{eqnarray}
Using that $\delta _{\lambda}A_t=t\delta _{\lambda} A+(1-t)\delta _{\lambda}\overline{A}=tD\lambda +(1-t)\overline{D}\lambda$
, the Bianchi identities $DF=0$ and $\overline{D}\overline{F}=0$, and the property of the covariant derivative and the invariant trace $<D(something)>=d<something>$ we get
\begin{eqnarray}
0=\delta _{\lambda}\mathcal{T}_{2n+1}=d\{(n+1)<F^n\lambda>-n(n+1)\int _0^1 dt~t<\Delta AF_t^{n-1}D\lambda >\}-\nonumber\\
-d\{ (n+1)<\overline{F}^{n}\lambda >-n(n+1)\int _0^1 dt~(t-1)<\Delta AF_t^{n-1}\overline{D}\lambda >\}
\end{eqnarray}
But the first line of the second member of the previous equation is just the gauge variation of the transgression if only $A$ 
is varied, which we will denote $\delta _{\lambda}\mathcal{T}^{(A)}_{2n+1}$. Therefore
\begin{eqnarray}
\delta _{\lambda}\mathcal{T}^{(A)}_{2n+1} =d\{ (n+1)<\overline{F}^{n}\lambda >-n(n+1)\int _0^1 dt~(t-1)<\Delta AF_t^{n-1}\overline{D}\lambda >\}\label{anomaly-backgrounds-1}
\end{eqnarray}
The expression within the brackets in the second member of the previous equation (upon which the exterior derivative $d$ acts) is already an expression of the consistent gauge anomaly in the presence of a background $\overline{A}$. By using that
$\overline{D}\lambda=d\lambda+[\overline{A},\lambda]$, that
$<\Delta AF_t^{n-1}d\lambda>=<d[\Delta AF_t^{n-1}]\lambda>-d<\Delta AF_t^{n-1}\lambda>$
and that $d^2=0$ we obtain
\begin{eqnarray}
\delta _{\lambda}\mathcal{T}^{(A)}_{2n+1} =d\{ (n+1)<\overline{F}^{n}\lambda >-n(n+1)\int _0^1 dt~(t-1)<d[\Delta AF_t^{n-1}]\lambda >-\nonumber\\-n(n+1)\int _0^1 dt~(t-1)<\Delta AF_t^{n-1}[\overline{A},\lambda ] >\}\label{anomaly-backgrounds-2}
\end{eqnarray} 
Eq.[\ref{anomaly-backgrounds-2}] differs from eq.[\ref{anomaly-backgrounds-1} ] in that the latter does not include $d\lambda$ while the former does.
If we write eq.[\ref{anomaly-backgrounds-2}] as $\delta _{\lambda}\mathcal{T}^{(A)}_{2n+1}=d\Omega ^1_{2n}(A,\overline{A},\lambda)$, which defines the anomaly 2n-form $\Omega ^1_{2n}$ in presence of a background, we get
\begin{eqnarray}
\Omega ^1_{2n}(A,\overline{A},\lambda)  =(n+1)<\overline{F}^{n}\lambda >-n(n+1)\int _0^1 dt~(t-1)<d[\Delta AF_t^{n-1}]\lambda >-\nonumber\\-n(n+1)\int _0^1 dt~(t-1)<\Delta AF_t^{n-1}[\overline{A},\lambda ] >\label{anomaly-backgrounds-3}
\end{eqnarray} 
In the particular case that $\overline{A}=0$ (therefore $\overline{F}=0$) eq.[\ref{anomaly-backgrounds-3}] reduces to
\begin{eqnarray}
\Omega ^1_{2n}(A,\overline{A}=0,\lambda)  = -n(n+1)\int _0^1 dt~(t-1)<d[AF_t^{n-1}]\lambda > \label{anomaly-wu-zee-zumino}
\end{eqnarray}
which agrees with the result Wu, Zee and Zumino for anomalies without backgrounds in Ref.\cite{wu-zee-zumino}(eq.(3.35) in that paper), if we take in account that they define gauge transformations as $\delta _{\lambda}A=-Dv$ (so we should replace $\lambda$ by $-v$).
Another particular case corresponds to choosing $\overline{A}=A$ the $\Delta A=0$ and
\begin{eqnarray}
\Omega ^1_{2n}(A,\overline{A}=A,\lambda)  =(n+1)<\overline{F}^{n}\lambda >\label{covariant-anomaly}
\end{eqnarray}  
which corresponds to the so called {\it covariant anomaly}.

\subsection{Consistent Anomaly and Covariant Anomaly}
 
If the {\it Quantum Effective Action}  \footnote{Also called {\it Quantum Action functional} and sometimes denoted $W[A]$ } of a gauge theory in a space-time of dimension $d$ is denoted $\Gamma [A]$ then its variation under infinitesimal gauge transformations with gauge parameter $\lambda$ defines the {\it Gauge Anomaly} d-form $\mathcal{A}[A,\lambda ]$  
\begin{eqnarray}
\delta _{\lambda}\Gamma [A]=\int _{\mathcal{M}^d}\mathcal{A}[A,\lambda ]
\end{eqnarray}
where ${\mathcal{M}^d}$ is the space-time manifold \cite{wu-zee-zumino}.

If the anomaly itself were the gauge variation of a local functional $\mathcal{F}[A]$ defined in ${\mathcal{M}^d}$, that is if $\mathcal{A}[A,\lambda ]=\delta _{\lambda}\mathcal{F}[A]$ then the anomaly can be  absorbed in a redefinition of the quantum action $\Gamma [A]\rightarrow \Gamma [A]-\mathcal{F}[A]$, and the anomaly for the new quantum action would be zero. 

The definition of the anomaly implies the so called {\it Wess-Zumino consistency condition} 
\begin{eqnarray}
\delta _{\eta}\mathcal{A}[A,\lambda ]-\delta _{\lambda}\mathcal{A}[A,\eta ]=\mathcal{A}[A,[\lambda ,\eta ] ]
\end{eqnarray}
were $\lambda$ and $\eta$ are infinitesimal gauge parameters. This condition severely restricts the possible form of the
 anomaly. The fact that $\delta _{\lambda}\mathcal{T}^{(A)}_{2n+1}=d\Omega ^1_{2n}(A,\overline{A},\lambda)$ implies that
$\Omega ^1_{2n}(A,\overline{A},\lambda)$ satisfies the Wess-Zumino consistency condition, assuming $\overline{A}$ is a fixed background that is not varied under gauge transformations. We have $\mathcal{A}[A,\lambda ]=\Omega ^1_{2n}(A,\overline{A},\lambda)$ as a possible consistent anomaly with background $\overline{A}$. That anomaly cannot be absorbed in a redefinition of the quantum action, as it is not the result of a variation of a local functional on $\mathcal{M}^d$ but rather is the result of the variation of a functional defined in a d+1-dimensional manifold with boundary $\mathcal{M}^d$. The previous paragraph parallels similar arguments for the case of anomalies without backgrounds (see \cite{wu-zee-zumino} and references therein).

The covariant anomaly eq.(\ref{covariant-anomaly}) is not consistent at 
first sight, as it is gauge invariant and therefore the first member of the 
consistency condition would vanish while the second member would not. 
However if we consider eq.(\ref{anomaly-backgrounds-3}) as the actual 
definition and $\overline{A}$ as a different field that is set equal 
to $A$ (for the covariant anomaly) only at the end of the calculations, 
and that is not to be varied under gauge transformations, then the 
covariant anomaly can also be regarded as consistent.

I believe that any proper definition of the anomaly must 
include the reference background (even if it were $A=0$).

\subsection{Diffeomorphism Anomalies with Backgrounds}

The non invariace of an action under diffeomorphisms is associated to gravitational 
and Weyl anomalies. The variation of an arbitrary differential form $\alpha _p$ 
under infinitesimal diffeomorphisms generated by the vector 
$\xi =\xi ^{\mu}\frac{\partial ~}{\partial x^{\mu}}$, taken at a fixed point, is $\delta _{\xi}\alpha _p=\mathcal{L}_{\xi}\alpha _p=[dI_{\xi}+I_{\xi}d]\alpha _p$, with $\mathcal{L}_{\xi}$ is the Lie derivative, $d$ is the standard exterior derivative and 
$I_{\xi }$ is the contraction operator
\footnote{The contraction operator $I_{\xi }$ is defined by
acting on a p-form $\alpha _p$ as
$$
I_{\xi }\alpha _{p}=\frac{1}{(p-1)!}\xi ^{\nu }\alpha _{\nu \mu
_{1}...\mu _{p-1}}dx^{\mu _{1}}...dx^{\mu _{p-1}}
$$
and being and anti-derivative in the sense that acting on the wedge
product of differential forms $\alpha _{p}$ and $\beta _{q}$ of
order $p$ and $q$ respectively gives $I_{\xi }(\alpha _{p}\beta
_{q})=I_{\xi }\alpha _{p}\beta _{q}+(-1)^{p}\alpha _{p}I_{\xi }\beta
_{q}$.} (see por instance \cite{nakahara}). For  a 1-form that is a gauge 
potential the previous expressions are equivalent to $\delta _{\xi}A=D[I_{\xi}A]+I_{\xi}F$ \footnote{And of course $\delta _{\xi}\overline{A}=\overline{D}[I_{\xi}\overline{A}]+I_{\xi}\overline{F}$.}.

Under infinitesimal diffeomorphisms the Transgression form changes as a 
differential form should $\delta _{\xi} \mathcal{T}_{2n+1} =\mathcal{L}_{\xi}\mathcal{T}_{2n+1}=[dI_{\xi}+I_{\xi}d]\mathcal{T}_{2n+1}$
(of course, in a $2n+1$ dimensional space-time $d\mathcal{T}_{2n+1}=0$, but we just kept that part as a formal relationship, that would be relevant if said space-time were embedded in a higher dimensional one). 
The variation of the Transgression results from variations from both $A$
and $\overline{A}$, and we could denote it as 
$\delta _{\xi} \mathcal{T}_{2n+1} =\delta _{\xi} \mathcal{T}_{2n+1}^{(A)}+\delta _{\xi} 
\mathcal{T}_{2n+1}^{(\overline{A})}$, where the first term of the second member comes from taking the variation of $A$ keeping $\overline{A}$ fixed and the second term to the reverse. 

If we understand $\overline{A}$ as a fixed background not to be varied and take variations only of $A$, then the consequent variation of the transgression will have two parts: a normal variation as an ordinary differential form under diffeomorphisms plus an anomalous variation. Explicitly 
$ \delta _{\xi} \mathcal{T}_{2n+1}^{(A)}=\delta _{\xi} \mathcal{T}_{2n+1}-\delta _{\xi} 
\mathcal{T}_{2n+1}^{(\overline{A})}$, where $\delta _{\xi} \mathcal{T}_{2n+1}$ is the normal variation and 
$-\delta _{\xi} \mathcal{T}_{2n+1}^{(\overline{A})}$ is the anomalous variation. That 
variation can be written as
\begin{eqnarray}
-\delta _{\xi} \mathcal{T}_{2n+1}^{(\overline{A})}=
I_{\xi}<\overline{F}^{n+1}>+d\{(n+1)<\overline{F}^n  I_{\xi}\overline{A}>-\nonumber\\
-n(n+1)\int _0^1dt(t-1)<\Delta AF_t^{n-1}I_{\xi}\overline{F}> -\nonumber\\
-n(n+1)\int _0^1dt(t-1)<\Delta AF_t^{n-1}\overline{D}I_{\xi}\overline{A}> 
\}\label{anomalous-variation-transgression-diffeos}
\end{eqnarray}
or
\begin{eqnarray}
-\delta _{\xi} \mathcal{T}_{2n+1}^{(\overline{A})}=
I_{\xi}<\overline{F}^{n+1}>+d \Omega ^1_{2n}(A,\overline{A},\xi)\label{anomalous-variation-transgression-diffeos-9}
\end{eqnarray}
with
\begin{eqnarray}
\Omega ^1_{2n}(A,\overline{A},\xi) =(n+1)<\overline{F}^n  I_{\xi}\overline{A}>-\nonumber\\
-n(n+1)\int _0^1dt(t-1)<\Delta AF_t^{n-1}I_{\xi}\overline{F}> -\nonumber\\
-n(n+1)\int _0^1dt(t-1)<\Delta AF_t^{n-1}\overline{D}I_{\xi}\overline{A}> 
\label{diffeo-anomaly-backgrounds}
\end{eqnarray}
A judicious choice of the background will kill the bulk term $I_{\xi}<\overline{F}^{n+1}>$, the simplest being $\overline{F}=0$, but we will see below 
other choices better suited for a well defined action principle for Chern-Simons AdS gravities. In that case 
the diffeomorphism anomaly with backgrounds is just $\Omega ^1_{2n}(A,\overline{A},\xi)$.

\section{Brief review of Transgression and Chern-Simons AdS gravity}

For the AdS group in dimension $d=2n+1$ the gauge connection is given by\footnote{A gauge
connection has dimensions of $(length)^{-1}$, so it must be  
$A=\frac{\omega ^{ab}}{2}J_{ab}+\frac{e^a}{l}P_a$ where $l$ is the 'AdS radius'.
I set $l=1$ trough all the present paper. It is
easy to reintroduce $l$ using dimensional
analysis, if necessary.}
$A=\frac{\omega ^{ab}}{2}J_{ab}+e^aP_a$ where $\omega ^{ab}$ is the spin connection, $e^a$ is the vielbein and $J_{ab}$ and $P_a$
are the generators of the AdS group (for Lorentz transformations and
translations respectively).  One possible symmetrized trace, and the only one I will consider in this paper, is that which is non zero only for one $P$ generator and $n$  $J$ generators, with values
\begin{equation}
<J_{a_{1}a_{2}}...J_{a_{2n-1}a_{2n}}P_{a_{2n+1}}>=\kappa\frac{2^{n}}
{(n+1)}\epsilon_{a_{1}...a_{2n+1}}
\end{equation}
where $\kappa$ is a constant, which together with the AdS group parameter $l$ ("AdS radius") will characterize the theories.
In addition to the basis of the algebra spanned by the generators $P_a$ and $J_{ab}$ we will
use a basis spanned by the generators $P_1$, $P_i$, $P_i+J_{1i}$ and $P_i-J_{1i}$, 
with $i$ an index taking any allowed value but 1. For this generators the only non 
zero values of the symmetrized trace are 
\begin{eqnarray}
<J_{i_{1}i_{2}}...J_{i_{2n-1}i_{2n}}P_{1}>=\kappa\frac{2^{n}}
{(n+1)}\epsilon_{1i_{1}...i_{2n}}\\
<J_{i_{1}i_{2}}...J_{i_{2n-1}i_{2n-2}}(P_{i_{2n-1}}\pm J_{1i_{2n-1}})
(P_{i_{2n}}\mp J_{1i_{2n}})>=\pm\kappa\frac{2^{n+1}}
{(n+1)}\epsilon_{1i_{1}...i_{2n}}
\end{eqnarray}
Notice in particular that
\begin{eqnarray}
<J_{i_{1}i_{2}}...J_{i_{2n-1}i_{2n-2}}(P_{i_{2n-1}}\pm J_{1i_{2n-1}})
(P_{i_{2n}}\pm J_{1i_{2n}})>= 0
\end{eqnarray}
The transgression for the AdS group  is\footnote{In what follows I will use a
compact notation where $\epsilon$ stands for the Levi-Civita symbol
$\epsilon _{a_1...a_d}$ and wedge products of differential forms are
understood, as it was done in Refs.\cite{motz2,motz1,motz3}. For instance: $\epsilon Re^{d-2}\equiv \epsilon
_{a_1a_2....a_d}R^{a_1a_2}\wedge e^{a_3}\wedge ...\wedge
e^{a_{d-2}}$,  $(\theta ^2)^{ab}=\theta ^a_c\wedge\theta ^{cb}$.} \cite{motz3}
\begin{equation}
{\cal T}_{2n+1}= \kappa \int _0^1dt \epsilon (R+t^2e^2)^ne -\kappa
\int _0^1dt \epsilon (\overline{R}+t^2\overline{e}^2)^n\overline{e}
+d~B _{2n}
\end{equation}
where
\begin{equation}
B _{2n}=-\kappa n\int_0^1dt\int_0^1ds~\epsilon\theta e_t
\left\{t R+(1-t)\overline{R}-t(1-t)\theta ^2+s^2e_t^2 \right\}^{n-1}
\end{equation}
Here $e ^a$ and $\overline{e}^a$ are the two vielbeins and $\omega
^{ab} $ and $\overline{\omega}^{ab}$ the two spin connections,
$R=d\omega +\omega ^2$ and
$\overline{R}=d\overline{\omega}+\overline{\omega}^2$ are the
corresponding curvatures, $\theta =\omega -\overline{\omega}$ and
$e_t=te+(1-t)\overline{e}$.  

The action for transgressions for the AdS group is chosen to be
\cite{motz3}
\begin{equation}
I_{Trans}= \kappa \int _{\cal M}\int _0^1dt \epsilon (R+t^2e^2)^ne
-\kappa\int_{\overline{{\cal M}}} \int _0^1dt \epsilon
(\overline{R}+t^2\overline{e}^2)^n\overline{e} +\int _{\partial {\cal
M} }B _{2n}\label{action-trangresion-ads}
\end{equation}
where ${\cal M}$ and $\overline{\cal M}$ are two manifolds with a
common boundary, that is $\partial {\cal M}\equiv\partial
\overline{{\cal M}}$. Notice that, as said in the previous section, this is a generalization from the
simpler case where ${\cal M}\equiv\overline{\cal M}$.  

We have two natural choices of  either regarding both $A$  and $\overline{A}$ as 
dynamical fields, or regarding one of them (lets say $\overline{A}$) as a 
non independent background.  

The field equations derived from the action of eq.(\ref{action-trangresion-ads}) are 
$<F^nG_{\alpha}>=0$ and $<\overline{F}^nG_{\alpha}>=0$, or (see for instance \cite{motz3}) 
\begin{eqnarray}
\epsilon (R+e^2)^n=0~~~,~~~\epsilon (R+e^2)^{n-1}T=0 \\
\epsilon (\overline{R}+\overline{e}^2)^n=0~~~,\label{field-equations-transgression-ads}
~~~\epsilon (\overline{R}+\overline{e}^2)^{n-1}\overline{T}=0
\end{eqnarray}
If $\overline{A}$ is taken to be non dynamical only the first line of the previous equations should hold.

What will be done in the next sections can be interpreted in two ways:

i. As the variation of the Transgression-Chern-Simons AdS gravity action (in any of its versions) under those gauge transformations that keep the AdS gauge curvature finite, and when $\overline{A}$ is not varied. 
In that sense the result can be regarded as the AdS gauge holographic anomaly for that theory.

ii. As the construction of the consistent gauge anomaly for the AdS group and the invariant tensor given above,
useful in principle (with a suitable coefficient) for other theories.

\section{AdS gauge transformations} 

In this section we study the generic asymptotic form of the gauge parameters that would generate gauge transformations consistent with the asymptotic gonditions in the gauge fields required in \cite{mora-action-principle-cs-ads}. A similar problem
has been considered, for different asymptotic conditions, in \cite{banados-miskovic-theisen}, and more recently in 
\cite{afshar} in 2+1 dimensions.
 
\subsection{Gauge transformation of the gauge potential}

Given the AdS gauge parameter $\lambda =\frac{1}{2}\lambda ^{ab}J_{ab}+\lambda ^aP_a$, the gauge potential
$A =\frac{1}{2}\omega ^{ab}J_{ab}+e ^aP_a$, the gauge variation $\delta _{\lambda}A=D\lambda =dA+[A,\lambda]$ and 
the algebra of generators of the AdS group one gets
\begin{eqnarray}
\delta _{\lambda}A=\frac{1}{2}[(d\lambda ^{ab}+\omega ^a_{~c}\lambda ^{cb}+\omega ^b_{~c}\lambda ^{ac})
+(e^a\lambda ^b - e^b\lambda ^a )]J_{ab} +\nonumber\\
+[ (d\lambda ^a+\omega ^a _{~b}\lambda ^b)-\lambda ^ a _{~b}e^b ]P_a
\end{eqnarray}
or 
\begin{eqnarray}
\delta _{\lambda}\omega ^{ab}=(d\lambda ^{ab}+\omega ^a_{~c}\lambda ^{cb}+\omega ^b_{~c}\lambda ^{ac})
+(e^a\lambda ^b - e^b\lambda ^a )= D\lambda ^{ab} + e^a\lambda ^b - e^b\lambda ^a    \nonumber\\
\delta _{\lambda}e^a =(d\lambda ^a+\omega ^a _{~b}\lambda ^b)-\lambda ^ a _{~b}e^b =
D\lambda ^a-\lambda ^ a _{~b}e^b 
\end{eqnarray}
where $D\lambda ^{ab}\equiv d\lambda ^{ab}+\omega ^a_{~c}\lambda ^{cb}+\omega ^b_{~c}\lambda ^{ac}$ and
$D\lambda ^a\equiv d\lambda ^a+\omega ^a _{~b}\lambda ^b$ are the 
covariant derivatives associated to the spin connection $\omega ^{ab}$.

In ref.\cite{mora-action-principle-cs-ads} it was shown that the gauge potential that results from requiring a finite gauge curvature can be cast in the form 
\begin{eqnarray}
A=e^r\hat{e}^i_{\infty}(P_i-J_{1i})+\nonumber\\
+\frac{1}{2}\hat{\omega}_{\infty}^{ij}J_{ij}+\frac{1}{2}\hat{\tau}^i(P_i-J_{1i})+dr~P_1+\nonumber\\
+\frac{1}{2}e ^{-r}[\hat{\gamma}^{ij}+dr\hat{\omega}_r^{ij}]J_{ij}-\frac{1}{2}e ^{-r}\hat{\zeta}^i(P_i+J_{1i})\label{asymptotic-A}
\end{eqnarray}
where we are using the notation of ref.\cite{mora-action-principle-cs-ads}. It is important for what follows to remark that the coefficient of $P_1$ does not need to be $dr$, but it is enough that it is finite when $r\rightarrow\infty $ in order to
yield a finite gauge curvature $F$ \footnote{The asymptotic dependence of $A$ on $r$ must be schematically 
$A\approx e^r(P_i-J_{1i})+ 1~J_{ij}+ 1 ~(P_i-J_{1i})+1~P_1+
e ^{-r}J_{ij}+e ^{-r}(P_i+J_{1i})$.}. 

We are interested in gauge transformations that preserve the finite character of $F$. A generic gauge transformation acting 
on $A$ of the form of eq.(\ref{asymptotic-A}) yields
\begin{eqnarray}
\delta _{\lambda} A=\frac{1}{2}\delta _{\lambda}\omega ^{ij} J_{ij}+\delta _{\lambda }e^1P_1+\nonumber\\
+\{ \frac{1}{2}\hat{D}_{\infty}\lambda ^{(-)i}-e^r\lambda ^{(-)i}_{~~~~j}\hat{e}_{\infty} ^j
-\frac{1}{2} \lambda ^{(-)i}_{~~~~j}\hat{\tau } ^j+ \nonumber\\
\frac{dr}{2}\left[  \partial _r\lambda ^{(-)i}- \lambda ^{(-)i}\right]  
+\frac{e^{-r}}{2}\left[ \hat{\gamma}^{i}_{~j}+dr\hat{\omega}_{r~j}^{i}\right]\lambda ^{(-)j} \}(P_i-J_{1i})\nonumber\\
+\{ \frac{1}{2}\hat{D}_{\infty}\lambda ^{(+)i}
+\frac{e^{-r}}{2} \lambda ^{(+)i}_{~~~~j}\hat{\zeta } ^j+ \nonumber\\
\frac{dr}{2}\left[  \partial _r\lambda ^{(+)i}+\lambda ^{(+)i}\right]  
+\frac{e^{-r}}{2}\left[ \hat{\gamma}^{i}_{~j}+dr\hat{\omega}_{r~j}^{i}\right]\lambda ^{(+)j} \}(P_i+J_{1i})
\end{eqnarray}
where  $\lambda ^{(\pm)i}=\lambda ^i\pm \lambda ^{1i}$ and 
$\lambda ^{(\pm)i}_{~~~~j}=\lambda ^i_{~j}\pm \lambda ^1\delta ^i_{~j}$, and 
\begin{eqnarray}
\delta _{\lambda}\omega ^{ij}= \hat{D} _{\infty }\lambda ^{ij}+dr\partial _r\lambda ^{ij}
+e^r[\hat{e}_{\infty} ^i \lambda ^{(+)j}-\lambda ^{(+)i}\hat{e}_{\infty} ^j]
+ \frac{1}{2}[\hat{\tau} ^i \lambda ^{(+)j}-\lambda ^{(+)i}\hat{\tau} ^j] +\nonumber\\
+ \frac{e^{-r}}{2}[\hat{\zeta} ^i \lambda ^{(-)j}-\lambda ^{(-)i}\hat{\zeta} ^j]  +
e^{-r}[\hat{\gamma}^{i}_{~k}+dr\hat{\omega}^{i}_{r~k}] \lambda ^{kj} +
e^{-r}[\hat{\gamma}^{j}_{~k}+dr\hat{\omega}^{j}_{r~k}] \lambda ^{ik} +\\
\delta _{\lambda}e^1=d\lambda ^1-e^r \lambda ^{(+)} _i\hat{e}_{\infty} ^i
-\frac{1}{2}\lambda ^{(+)}_i\hat{\tau} ^i-\frac{e^{-r}}{2}\lambda ^{(-)}_i\hat{\zeta} ^i
\end{eqnarray}

\subsection{Gauge transformations of the gauge curvature}

The gauge variation of the AdS gauge curvature is $\delta _{\lambda}F=[F,\lambda ]$. This implies that the
components of $F$ transform as 
\begin{eqnarray}
\delta _{\lambda}F^a=F^a_{~b}\lambda ^b-\lambda ^a_{~b}F^b\nonumber\\
\delta _{\lambda}F^{ab}=-\lambda^{a}_{~c}F^{cb}-\lambda ^b_{~c}F^{ac}+F^a\lambda ^b -F^b\lambda ^a\label{gauge variation-F}
\end{eqnarray}
Separating the $1$ and the $i\neq 1$ components we get
\begin{eqnarray}
\delta _{\lambda}F^i=F^i_{~j}\lambda ^j + F^i_{~1}\lambda ^1  -\lambda ^i_{~j}F^j-\lambda ^i_{~1}F^1\nonumber\\
\delta _{\lambda}F^1=F^1_{~i}\lambda ^i-\lambda ^1_{~i}F^i\nonumber\\
\delta _{\lambda}F^{ij}=-\lambda^{i}_{~k}F^{kj}-\lambda ^j_{~k}F^{ik}-\lambda^{i}_{~1}F^{1j}-\lambda ^j_{~1}F^{i1}
+F^i\lambda ^j -F^j\lambda ^i  \nonumber\\
\delta _{\lambda}F^{1i}=-\lambda^{1}_{~j}F^{ji}-\lambda ^i_{~j}F^{1j}+F^1\lambda ^i -F^i\lambda ^1
\end{eqnarray}
or equivalently
\begin{eqnarray}
\delta _{\lambda}F^i=\frac{1}{2} F^{(+)i}_{~~~~j}\lambda ^{(+)j} + \frac{1}{2} F^{(-)i}_{~~~~j}\lambda ^{(-)j} + F^i_{~1}\lambda ^1  -\lambda ^i_{~j}F^j\nonumber\\
\delta _{\lambda}F^1=\frac{1}{2}F^{(+)}_{~i}\lambda ^{(-)i}- \frac{1}{2}F^{(-)}_{~i}\lambda ^{(+)i}   \nonumber\\
\delta _{\lambda}F^{ij}=-\lambda^{i}_{~k}F^{kj}-\lambda ^j_{~k}F^{ik}-\nonumber\\
-\frac{1}{2}\lambda^{(+)i}F^{(-)j}-\frac{1}{2}\lambda ^{(-)i}F^{(+)j}
+\frac{1}{2}\lambda^{(+)j}F^{(-)i}+\frac{1}{2}\lambda ^{(-)j}F^{(+)i} \nonumber\\
\delta _{\lambda}F^{1i}=\frac{1}{2}\lambda^{(-)j}F_j^{(+)i}-\frac{1}{2}\lambda^{(+)j}F_j^{(-)i}
-\lambda ^i_{~j}F^{1j} -F^i\lambda ^1
\end{eqnarray}
where  $F ^{(\pm)i}=F ^i\pm F ^{1i}$ and 
$F ^{(\pm)i}_{~~~~j}=F ^i_{~j}\pm F ^1\delta ^i_{~j}$.

\subsection{Asymptotic conditions on the gauge parameters}

We must choose what conditions to impose on the asymptotic dependence of the components of $\lambda $ on $r$. 
That may seem unnecessary, as 
$\lambda $ corresponds to an infinitesimal gauge transformation. However we may regard $\lambda $ as infinitesimal
at any given large but finite $r$, but with an asymptotic dependence that would render it infinite if 
$r\rightarrow \infty $ and everything else is kept fixed. Considering that the components of $\lambda$ are of the generic form 
$\sigma f(x,r)$ where $\sigma $ is some infinitesimal parameter, we must deal with two different limits
$\sigma \rightarrow 0$ (although never in fact reaching 0) and $r\rightarrow\infty$.

I see at least two possible conditions on the asymptotic behavior of $\lambda$:\\

{\bf Option I.} The first possibility is to require that gauge parameter must be such that 
the infinitesimal gauge transformation preserves 
the generic asymptotic form of $A$ that yields a finite asymptotic $F$. 
In that case, from the expressions 
given above for the gauge variation of $A$ and $F$, we see that we must require:\\
i. Requiring that $e^1$ and $\omega ^{ij}$ are kept asymptotically 
finite by the allowed gauge transformations implies that the functions  
$\lambda ^{(+)i}(x,r)$ are asymptotically of the form
$\lambda ^{(+)i}(x,r)= e^{-r}\hat{\lambda } ^{(+)i}(x,r)$ with 
$\hat{\lambda } ^{(+)i}(x,r)$ asymptotically finite.\\
ii. The functions $\lambda ^{(-)i}(x,r)$,  $\lambda ^{ij}(x,r)$ and $\lambda ^1(x,r)$ 
(and equivalently  
$\lambda ^{(+)ij}(x,r)$ and $\lambda ^{(-)ij}(x,r)$) are asymptotically finite. 

Allowing $\lambda ^{(-)i}(x,r) = e^r\lambda ^{(-)i}(x,r)$ with $\lambda ^{(-)i}(x,r)$
asymptotically finite would preserve the generic asymptotic behaviour of $A$, but
would make some components of $F$ asymptotically divergent, therefore we forbid that 
possibility.\\

We have then
\begin{eqnarray}
\delta _{\lambda} A=\frac{1}{2}\delta _{\lambda}\omega ^{ij} J_{ij}+\delta _{\lambda }e^1P_1+\nonumber\\
+\{ e^r [-\lambda ^{(-)i}_{~~~~j}\hat{e}_{\infty} ^j
 ]
+\frac{dr}{2}[ \partial _r \lambda  ^{(-)i} - \lambda  ^{(-)i} ]+\frac{1}{2}\hat{D}_{\infty} \lambda ^{(-)i}-\frac{1}{2} \lambda ^{(-)i}_{~~~~j}\hat{\tau } ^j+\nonumber\\
+\frac{e^{-r}}{2}\left[ \hat{\gamma}^{i}_{~j}+dr\hat{\omega}_{r~j}^{i}\right]\lambda ^{(-)j} \}(P_i-J_{1i})\nonumber\\
+\{ \frac{e^{-r}}{2}\left[\hat{D}_{\infty}\hat{\lambda } ^{(+)i}
+ \lambda ^{(+)i}_{~~~~j}\hat{\zeta } ^j+ 
dr \partial _r\hat{\lambda }^{(+)i} \right]
+\frac{e^{-2r}}{2}\left[ \hat{\gamma}^{i}_{~j}+dr\hat{\omega}_{r~j}^{i}\right]\hat{\lambda } ^{(+)j} \}(P_i+J_{1i})\label{asymptotic-gauge-transformations-A-I}
\end{eqnarray}
and
\begin{eqnarray}
\delta _{\lambda}\omega ^{ij}= \hat{D} _{\infty }\lambda ^{ij}+dr\partial _r\lambda ^{ij}
+[\hat{e}_{\infty} ^i \hat{\lambda } ^{(+)j}-\hat{\lambda }^{(+)i}\hat{e}_{\infty} ^j]
+ \frac{e^{-r}}{2}[\hat{\zeta} ^i \lambda ^{(-)j}-\lambda ^{(-)i}\hat{\zeta} ^j] +\nonumber\\
+\frac{e^{-r}}{2}[\hat{\tau} ^i \hat{\lambda }^{(+)j}-\hat{\lambda }^{(+)i}\hat{\tau} ^j] +
e^{-r}[\hat{\gamma}^{i}_{~k}+dr\hat{\omega}^{i}_{r~k}] \lambda ^{kj} +
e^{-r}[\hat{\gamma}^{j}_{~k}+dr\hat{\omega}^{j}_{r~k}] \lambda ^{ik} \label{asymptotic-gauge-transformation-omega-I}\\
\delta _{\lambda} e^1=d\lambda ^1- \hat{\lambda } ^{(+)} _i\hat{e}_{\infty} ^i
-\frac{e^{-r}}{2}\lambda  ^{(-)}_i\hat{\zeta} ^i
-\frac{e^{-r}}{2}\hat{\lambda } ^{(+)}_i\hat{\tau} ^i\label{asymptotic-gauge-transformations-e1-I}
\end{eqnarray}
From the previous expressions we can read the variations of the different relevant fields, for instance
\begin{eqnarray}
\delta _{\lambda}\hat{e}_{\infty} ^i= -\lambda ^{(-)i}_{\infty ~~j}\hat{e}_{\infty} ^j \label{asymptotic-gauge-transformations-boundary-vielbein-I}
\end{eqnarray}
where   $\lambda ^{(-)i}_{\infty ~~j}(x)=\lambda ^{(-)i}_{~~~~j}(x, r\rightarrow\infty )$.  

If $\lambda ^1$ is only function of the $x$'s but not of $r$ asymptotically, or more precisely if $\partial _r\lambda ^1\rightarrow 0 $ when $r\rightarrow\infty$, we can make $\delta _{\lambda} e^1 =0$ asymptotically
by choosing the $\hat{\lambda } ^{(+)} _i(x,r\rightarrow\infty )$ to be the components of $d\lambda ^1$ in the basis
$\hat{e}_{\infty} ^i$ \footnote{Whether or not we make this choice will not affect the anomalies computed below.}. 
Making $\delta _{\lambda} e^1 =0$ everywhere would require gauge parameters that are dependent on the specific field configuration. We will not make that sort of choice. If $\delta _{\lambda} e^1 =0$ at the boundary the gauge potential could in principle be transformed to the form with which we started, with  $e^1=dr$, by a change of coordinates that reduces to the identity at the boundary.\\

{\bf Option II.} The second possibility is to require that the asymptotic behavior of the gauge parameters is
such that that the gauge Noether's charge densities are finite, which may be achieved but requiring that it is 
the same than the asymptotic behavior of $I_{\xi}A$. 

From the invariance of the action under diffeomorphisms generated by a infinitesimal space-time vector $\xi ^{\mu}$ 
Noether's Theorem yields (see for instance \cite{motz3}) the conserved current
\begin{equation}
\ast j_{\xi} = ~d Q _{\xi}
\end{equation}
where the conserved charge density is
\begin{eqnarray}
Q _{\xi} = +n(n+1)\int_0^1dt <\Delta A F_t^{n-1} I _{\xi }A _t>\label{Noether-charge-diffeo}
\end{eqnarray}
Analogously,  invariance of the action under gauge transformations generated by the algebra-valued gauge parameter 
$\lambda$ gives, via
Noether's Theorem \cite{motz3}, the conserved current
\begin{equation}
\ast j_{\lambda} = ~d Q _{\lambda}
\end{equation}
where the conserved charge density is
\begin{eqnarray}
Q _{\lambda} = +n(n+1)\int_0^1dt <\Delta A F_t^{n-1} \lambda >\label{Noether-charge-gauge}
\end{eqnarray}
We see that bot expressions are the same if we replace $I _{\xi }A _t$ by $\lambda$. Therefore, using the results of ref.\cite{mora-action-principle-cs-ads}, were it was shown that the asymptotic conditions given ensure a finite $Q _{\xi}$, if the asymptotic behavior of $\lambda$ is the same than $I _{\xi }A _t$ then $Q _{\lambda}$ will be finite. Reasoning as in 
ref.\cite{mora-action-principle-cs-ads} one can see that the asymptotic behavior of Option I gives vanishing 
$Q _{\lambda}$'s.

This weaker condition implies the asymptotic dependence 
on $A$ is preserved, but also that some components of $F$ would diverge if we take the $r\rightarrow\infty$ 
while keeping $\sigma $ (the infinitesimal factor mentioned above) fixed. In fact, as we are considering infinitesimal transformations, one must regard $\sigma$ as small enough for a given $r$ that the gauge variation of $F$ is in fact infinitesimal (therefore keeping it finite). 

From the expressions for  $A$ and for the gauge variation of $A$ we see that we must require:\\
i. The functions $\lambda ^{ij}(x,r)$ and $\lambda ^1(x,r)$ and equivalently  
$\lambda ^{(+)ij}(x,r)$ and $\lambda ^{(-)ij}(x,r)$ are asymptotically finite.\\
ii. The functions  $\lambda ^{(+)i}(x,r)$ are asymptotically of the form
$\lambda ^{(+)i}(x,r)= e^{-r}\hat{\lambda } ^{(+)i}(x,r)$ with 
$\hat{\lambda } ^{(+)i}(x,r)$ asymptotically finite.\\
iii. The functions  $\lambda ^{(-)i}(x,r)$ are asymptotically of the form
$\lambda ^{(-)i}(x,r)= e^{r}\hat{\lambda } ^{(-)i}(x,r)$ with 
$\hat{\lambda } ^{(+)i}(x,r)$ asymptotically finite.\\

We have then
\begin{eqnarray}
\delta _{\lambda} A=\frac{1}{2}\delta _{\lambda}\omega ^{ij} J_{ij}+\delta _{\lambda }e^1P_1+\nonumber\\
+\{ e^r [\frac{1}{2}\hat{D}_{\infty}\hat{\lambda } ^{(-)i}-\lambda ^{(-)i}_{~~~~j}\hat{e}_{\infty} ^j
+\frac{dr}{2} \partial _r\hat{\lambda } ^{(-)i}  ]-\nonumber\\
-\frac{1}{2} \lambda ^{(-)i}_{~~~~j}\hat{\tau } ^j
+\frac{1}{2}\left[ \hat{\gamma}^{i}_{~j}+dr\hat{\omega}_{r~j}^{i}\right]\hat{\lambda }^{(-)j} \}(P_i-J_{1i})\nonumber\\
+\{ \frac{e^{-r}}{2}\left[\hat{D}_{\infty}\hat{\lambda } ^{(+)i}
+ \lambda ^{(+)i}_{~~~~j}\hat{\zeta } ^j+ 
dr \partial _r\hat{\lambda }^{(+)i} \right]
+\frac{e^{-2r}}{2}\left[ \hat{\gamma}^{i}_{~j}+dr\hat{\omega}_{r~j}^{i}\right]\hat{\lambda } ^{(+)j} \}(P_i+J_{1i})\label{asymptotic-gauge-transformations-A-II}
\end{eqnarray}
and
\begin{eqnarray}
\delta _{\lambda}\omega ^{ij}= \hat{D} _{\infty }\lambda ^{ij}+dr\partial _r\lambda ^{ij}
+[\hat{e}_{\infty} ^i \hat{\lambda } ^{(+)j}-\hat{\lambda }^{(+)i}\hat{e}_{\infty} ^j]
+ \frac{1}{2}[\hat{\zeta} ^i \hat{\lambda }^{(-)j}-\hat{\lambda }^{(-)i}\hat{\zeta} ^j] +\nonumber\\
+\frac{e^{-r}}{2}[\hat{\tau} ^i \hat{\lambda }^{(+)j}-\hat{\lambda }^{(+)i}\hat{\tau} ^j] +
e^{-r}[\hat{\gamma}^{i}_{~k}+dr\hat{\omega}^{i}_{r~k}] \lambda ^{kj} +
e^{-r}[\hat{\gamma}^{j}_{~k}+dr\hat{\omega}^{j}_{r~k}] \lambda ^{ik} \label{asymptotic-gauge-transformation-omega-II}\\
\delta _{\lambda} e^1=d\lambda ^1- \hat{\lambda } ^{(+)} _i\hat{e}_{\infty} ^i
-\frac{1}{2}\hat{\lambda } ^{(-)}_i\hat{\zeta} ^i
-\frac{e^{-r}}{2}\hat{\lambda } ^{(+)}_i\hat{\tau} ^i\label{asymptotic-gauge-transformations-e1-II}
\end{eqnarray}
From the previous expressions we can read the variations of the different relevant fields, for instance
\begin{eqnarray}
\delta _{\lambda}\hat{e}_{\infty} ^i=\frac{1}{2}\hat{D}_{\infty}\hat{\lambda } ^{(-)i}-\lambda ^{(-)i}_{~~~~j}\hat{e}_{\infty} ^j
+\frac{dr}{2} \partial _r\hat{\lambda } ^{(-)i} \label{asymptotic-gauge-transformations-boundary-vielbein-II}
\end{eqnarray}
where every dependence on $r$ is taken on the limit $r\rightarrow\infty$. If we require that $\delta _{\lambda}\hat{e}_{\infty} ^i$ has no component along $dr$ we must 
require that $\partial _r\hat{\lambda } ^{(-)i}\rightarrow 0$ when $r\rightarrow \infty $.

Making  $\delta _{\lambda} e^1 =0$ at the boundary is less straightforward in this case, 
as its finite part involves the specific field configuration considered, through $\hat{\zeta} ^i$, then 
the only configuration independent choice would be $\hat{\lambda } ^{(-)}_i\rightarrow 0$ asymptotically, which in fact corresponds to Option I. It seems that in this case
 we must regard the asymptotic vanishing of $\delta _{\lambda} e^1 =0$
as a configuration dependent equation that yields configuration dependent asymptotic 
allowed values for $\hat{\lambda } ^{(+)i}$ 
and $\hat{\lambda } ^{(-)i}$
\begin{eqnarray}
d\lambda ^1- \hat{\lambda } ^{(+)} _i\hat{e}_{\infty} ^i
-\frac{1}{2}\hat{\lambda } ^{(-)}_i \hat{\zeta} ^i=0
\end{eqnarray}

\subsubsection{Discussion}

Option I above is somehow "safer", as $F$ will be kept finite independently 
of the order in which we take limits, and also allows for a simple configuration 
independent condition to make $\delta _{\lambda}e^1=0$. It has the disadvantage that 
the Noether's charge associated to the gauge invariance does vanish in this case.

Option II is more general, containing all the gauge transformations allowed by Option I
plus others, and yields finite conserved charges consistent with the Noether's charges associated to 
diffeomorphisms. It has the problems that $\delta _{\lambda}F$ would diverge 
asymptotically if we are not careful about the order in which limits are taken.
Also the condition $\delta _{\lambda}e^1=0$ is more involved and configuration dependent in this case.

In the following sections on anomalies we will give the results corresponding to Option II. 
The anomalies corresponding to Option I are found by setting 
$\hat{\lambda } ^{(-)i}(x,r\rightarrow\infty )=\hat{\lambda } ^{(-)i}_{\infty}=0$ in the results
for Option II.
  
\section{AdS gauge Anomalies of Chern-Simons AdS gravity: Backgrounds}

\subsection{AdS gauge anomaly}

We will use the form of the anomaly given in eq.(\ref{anomaly-backgrounds-1}), in a 
slightly modified form
\begin{eqnarray}
\Omega ^{1}_{2n}(A,\overline{A},\lambda ) = (n+1)<\overline{F}^{n}\lambda >
-n(n+1)\int _0^1 dt~(t-1)<\Delta AF_t^{n-1}\delta _{\lambda }\overline{A} >\label{anomaly-backgrounds-9}
\end{eqnarray}
For the "AdS vacuum" configuration (see ref.\cite{mora-action-principle-cs-ads}) 
we have that $ <\overline{F}^{n}\lambda >=0 $, because it satisfies the classical field equations. We need to see which parts of $<\Delta AF_t^{n-1}\delta _{\lambda }\overline{A} >$ do not vanish as a result of the traces or the asymptotic behaviour of the fields. Schematically, the leading order in each generator is
\begin{eqnarray}
\Delta A\approx  e^{-r}~J_{ij}+ 1 ~(P_i-J_{1i})+e ^{-r}(P_i+J_{1i})\nonumber\\
F_t\approx  1~J_{ij}+ 1 ~(P_i-J_{1i})+1~P_1+e ^{-r}(P_i+J_{1i})\nonumber\\
\delta _{\lambda }\overline{A}\approx e^r(P_i-J_{1i})+ 1~J_{ij}+1~P_1+e ^{-r}(P_i+J_{1i})
\end{eqnarray}
Proceeding as in ref.\cite{mora-action-principle-cs-ads} we see that there are no divergences and that the only finite contribution comes from taking $\Delta A$ along $(P_i+J_{1i})$,
all the $F_t$'s along $J_{ij}$ and $\delta _{\lambda }\overline{A}$ along $(P_i-J_{1i})$. 
Using the definition of the symmetrized trace we get 
\begin{eqnarray}
\Omega ^{1}_{2n}(A,\overline{A},\lambda ) =\nonumber\\ 
=2 n \kappa  \int_0^1dt~(t-1) ~\epsilon _{ijk_1l_1...k_{n-1}l_{n-1}}
\Delta \hat{\zeta}^i [\frac{1}{2}\hat{D}_{\infty}\hat{\lambda } ^{(-)j} -\lambda ^{(-)j}_{~~~~k}\hat{e}_{\infty} ^k]\nonumber\\
(\hat{R}^{k_1l_1}_{\infty}-  
\hat{\zeta }_t^{k_1}\hat{e}^{l_1}_{\infty}-\hat{e}^{k_1}_{\infty}\hat{\zeta }_t^{l_1})
 ... 
(\hat{R}^{k_{n-1}l_{n-1}}_{\infty}-  
\hat{\zeta }_t^{k_{n-1}}\hat{e}^{l_{n-1}}_{\infty}-\hat{e}^{k_{n-1}}_{\infty}\hat{\zeta }_t^{l_{n-1}})  
\end{eqnarray}
The AdS vacuum is such that $\hat{R}^{ij}_{\infty}-  
\hat{\overline{\zeta }}^i\hat{e}^j_{\infty}-\hat{e}^i_{\infty}\hat{\overline{\zeta }}^j=0$, then 
$\hat{R}^{kl}_{\infty}-  
\hat{\zeta }_t^{k}\hat{e}^{l}_{\infty}-\hat{e}^{k}_{\infty}\hat{\zeta }_t^{l}=- t (
\Delta\hat{\zeta }^{k}\hat{e}^{l}_{\infty}+\hat{e}^{k}_{\infty}\Delta\hat{\zeta }^{l})$.
Using this in the previous expression we get
\begin{eqnarray}
\Omega ^{1}_{2n}(A,\overline{A},\lambda ) =\nonumber\\
= (-1)^{n}   2 n \kappa\int_0^1dt~(t^n-t^{n-1}) ~ ~\epsilon _{ijk_1l_1...k_{n-1}l_{n-1}}
\Delta \hat{\zeta}^i  [\lambda ^{(-)j}_{~~~~k}\hat{e}_{\infty} ^k-\frac{1}{2}\hat{D}_{\infty}\hat{\lambda } ^{(-)j} ] \nonumber\\
(\Delta\hat{\zeta }^{k_1}\hat{e}^{l_1}_{\infty}+\hat{e}^{k_1}_{\infty}\Delta\hat{\zeta }^{l_1})
 ... 
(\Delta\hat{\zeta }^{k_{n-1}}\hat{e}^{l_{n-1}}_{\infty}+\hat{e}^{k_{n-1}}_{\infty}\Delta\hat{\zeta }^{l_{n-1}})  
\end{eqnarray}
Integrating in the parameter $t$ we get
\begin{eqnarray}
\Omega ^{1}_{2n}(A,\overline{A},\lambda )= \frac{(-1)^{n+1}   2  \kappa}{n+1}  ~\epsilon _{ijk_1l_1...k_{n-1}l_{n-1}}
\Delta \hat{\zeta}^i  [  \lambda ^{(-)j}_{~~~~k}\hat{e}_{\infty} ^k-\frac{1}{2}\hat{D}_{\infty}\hat{\lambda } ^{(-)j} ] \nonumber\\
(\Delta\hat{\zeta }^{k_1}\hat{e}^{l_1}_{\infty}+\hat{e}^{k_1}_{\infty}\Delta\hat{\zeta }^{l_1})
 ... 
(\Delta\hat{\zeta }^{k_{n-1}}\hat{e}^{l_{n-1}}_{\infty}+\hat{e}^{k_{n-1}}_{\infty}\Delta\hat{\zeta }^{l_{n-1}})  
\end{eqnarray}
Using $\hat{R}^{ij}_{\infty}-  
\hat{\overline{\zeta }}^i\hat{e}^j_{\infty}-\hat{e}^i_{\infty}\hat{\overline{\zeta }}^j=0$ we can show that 
$\Delta\hat{\zeta }^{k}\hat{e}^{l}_{\infty}+\hat{e}^{k}_{\infty}\Delta\hat{\zeta }^{l}=
-F^{kl} (x,r=\infty)\equiv -F^{kl}_{\infty}$, which implies that the AdS gauge anomaly with backgrounds can be written as
\begin{eqnarray}
\Omega ^{1}_{2n}(A,\overline{A},\lambda )=\frac{   2  \kappa}{n+1}  ~\epsilon _{ijk_1l_1...k_{n-1}l_{n-1}}
\Delta \hat{\zeta}^i  [ \lambda ^{(-)j}_{~~~~k}\hat{e}_{\infty} ^k-\frac{1}{2}\hat{D}_{\infty}\hat{\lambda } ^{(-)j} ] 
F^{k_1l_1}_{\infty}
 ... 
F^{k_{n-1}l_{n-1}}_{\infty}  \label{AdS-anomaly-backgrounds}
\end{eqnarray}
It is important to remark that while we only used the subscript $\infty$ for
$ F^{kl}_{\infty} $, in fact every function in the previous expression is evaluated at the boundary $r=\infty$.

\subsection{Weyl anomaly}

From eq.(\ref{asymptotic-gauge-transformations-boundary-vielbein-I}) or 
eq.(\ref{asymptotic-gauge-transformations-boundary-vielbein-II}) we see that if the
only non zero component of the gauge parameter is $\lambda ^1$ then 
\begin{eqnarray}
\delta _{\lambda}\hat{e}_{\infty} ^i= \lambda ^{1}_{\infty}\hat{e}_{\infty} ^i \label{weyl-transformations-boundary-vielbein}
\end{eqnarray}
which implies that this kind of gauge transformation induces a Weyl transformation at the boundary. From eq.(\ref{AdS-anomaly-backgrounds}) we see that the  Weyl anomaly reads
\begin{eqnarray}
\Omega ^{1}_{2n}(A,\overline{A},\lambda )=-\frac{   2  \kappa \lambda ^1_{\infty}}{n+1}  ~\epsilon _{ijk_1l_1...k_{n-1}l_{n-1}}
\Delta \hat{\zeta}^i _{\infty}   \hat{e}_{\infty} ^j 
F^{k_1l_1}_{\infty}
 ... 
F^{k_{n-1}l_{n-1}}_{\infty}=\nonumber\\
= -\frac{    \kappa \lambda ^1_{\infty}}{n+1}  ~\epsilon _{ijk_1l_1...k_{n-1}l_{n-1}}
[\Delta \hat{\zeta}^i _{\infty}   \hat{e}_{\infty} ^j+ \hat{e}_{\infty} ^i \Delta \hat{\zeta}^j_{\infty}  ]
F^{k_1l_1}_{\infty}
 ... 
F^{k_{n-1}l_{n-1}}_{\infty}=\nonumber\\
= \frac{    \kappa \lambda ^1_{\infty}}{n+1}  ~\epsilon _{ijk_1l_1...k_{n-1}l_{n-1}}
F^{ij}_{\infty}F^{k_1l_1}_{\infty}
 ... 
F^{k_{n-1}l_{n-1}}_{\infty}\label{weyl-anomaly-backgrounds}
\end{eqnarray}
The last line would vanish if the field equations hold, but that does not mean that the Weyl anomaly vanishes, because gauge symmetries must hold whether or not the field equations hold (they hold off-shell, not just on-shell)\footnote{One may however argue that if there exist a holographic theory induced at the boundary which could be approximated by a saddle point approximation of the bulk theory, then the Weyl anomaly for that conjectured holographic boundary theory in that regime would vanish.}.

\subsection{Lorentz anomaly}

From eq.(\ref{asymptotic-gauge-transformations-boundary-vielbein-I}) or 
eq.(\ref{asymptotic-gauge-transformations-boundary-vielbein-II}) we see that if the
only non zero component of the gauge parameter is $\lambda ^{ij}$ then 
\begin{eqnarray}
\delta _{\lambda}\hat{e}_{\infty} ^i= -\lambda ^{i}_{\infty j}\hat{e}_{\infty} ^j \label{lorentz-transformations-boundary-vielbein}
\end{eqnarray}
which implies that this kind of gauge transformation induces a Lorentz transformation 
at the boundary. The corresponding anomaly can be interpreted as a 
Lorentz anomaly (which is equivalent to a gravitational anomaly, as shown in 
ref.\cite{bardeen-zumino} ). From eq.(\ref{AdS-anomaly-backgrounds}) we see that the Lorentz 
anomaly reads
\begin{eqnarray}
\Omega ^{1}_{2n}(A,\overline{A},\lambda )=\frac{   2  \kappa}{n+1}  ~\epsilon _{ijk_1l_1...k_{n-1}l_{n-1}}
\Delta \hat{\zeta}^i  [ \lambda ^{j}_{\infty k}\hat{e}_{\infty} ^k ] 
F^{k_1l_1}_{\infty}
 ... 
F^{k_{n-1}l_{n-1}}_{\infty}  \label{lorentz-anomaly-backgrounds}
\end{eqnarray}
This expression is not necessarily zero even if the field equations hold. It is
unclear to me which kind of gravitational coupling in a boundary theory may generate such Lorentz-gravitational anomaly (if any), as the usual gravitational anomalies are associated to symmetrized standard traces or products of standard traces, rather than to the symmetrized trace associated to the Levi-Civita tensor.

\subsection{Gauge translations anomaly}

If the only non zero components of the gauge parameter are $\hat{\lambda } ^{(-)i}$ we could in principle
have a gauge translation anomaly, which using eq.(\ref{AdS-anomaly-backgrounds}) would be
\begin{eqnarray}
\Omega ^{1}_{2n}(A,\overline{A},\lambda )=-\frac{    \kappa}{n+1}  ~\epsilon _{ijk_1l_1...k_{n-1}l_{n-1}}
\Delta \hat{\zeta}^i  [\hat{D}_{\infty}\hat{\lambda } ^{(-)j} ] 
F^{k_1l_1}_{\infty}
 ... 
F^{k_{n-1}l_{n-1}}_{\infty}\label{gauge-translation-anomaly-backgrounds}
\end{eqnarray}
Notice that for the Option I of the asymptotic behavior of the gauge parameter
the previous expression vanishes, and therefore in that case 
there are no anomalies associated to gauge translations 
in the backgrounds approach to Chern-simons AdS gravity.

\section{AdS gauge Anomalies of Chern-Simons AdS gravity: Kounterterms}

\subsection{AdS gauge anomaly}

We will again use the form of the anomaly given by eq.(\ref{anomaly-backgrounds-9}).

For the "Kounterterms vacuum" configuration we have 
$\overline{A}=\frac{1}{2}\hat{\omega}^{ij}_{\infty}J_{ij}+drP_1$ and
$\overline{F}= \frac{1}{2}\hat{R}^{ij}_{\infty}J_{ij}$
(see ref.\cite{mora-action-principle-cs-ads}), therefore $ <\overline{F}^{n}\lambda >$
has a non-zero contribution from the component of $\lambda $ along $P_1$, which is
$$
(n+1)<\overline{F}^{n}\lambda >= \kappa  ~\epsilon _{k_1l_1...k_{n}l_{n}}
\hat{R}^{k_1l_1}_{\infty} ... 
\hat{R}^{k_{n}l_{n}}_{\infty}  \lambda ^{1}=\kappa E_n\lambda ^1
$$
where $E_n=\epsilon _{k_1l_1...k_{n}l_{n}}\hat{R}^{k_1l_1}_{\infty} ... 
\hat{R}^{k_{n}l_{n}}_{\infty} $ is the Euler density of the boundary.

In order to compute the non zero contributions to the anomaly coming from $<\Delta AF_t^{n-1}\delta _{\lambda }\overline{A} >$ we need the asymptotic behaviour of the relevant fields, which schematically is
\begin{eqnarray}
\Delta A\approx  e^{-r}~J_{ij}+ e^r ~(P_i-J_{1i})+e ^{-r}(P_i+J_{1i})\nonumber\\
F_t\approx  1~J_{ij}+ 1 ~(P_i-J_{1i})+1~P_1+e ^{-r}(P_i+J_{1i})\nonumber\\
\delta _{\lambda }\overline{A}\approx 1~J_{ij}+e^r~(P_i-J_{1i})+ 1~P_1+e ^{-r}(P_i+J_{1i})
\end{eqnarray}
These asymptotic dependences come from the expressions for $\Delta A$ and $F_t$ given in
ref.\cite{mora-action-principle-cs-ads} and from 
\begin{eqnarray}
\delta _{\lambda }\overline{A}=\frac{1}{2}[\hat{D}_{\infty}\lambda ^{ij}
+dr\partial _r\lambda ^{ij}] J_{ij} + \frac{e^r}{2}[\hat{D}_{\infty}\hat{\lambda }^{(-)i}
+dr(\partial _r\hat{\lambda } ^{(-)i} +\hat{\lambda } ^{(-)i}   )](P_i-J_{1i}) +\nonumber\\
+ \frac{e^{-r}}{2}[\hat{D}_{\infty}\lambda ^{(+)i}
+dr(\partial _r\hat{\lambda }^{(+)i}- \hat{\lambda }^{(+)i} )](P_i+J_{1i})+
[\hat{d}\lambda ^{1}
+dr\partial _r\lambda ^{1}]P_1\label{gauge-variation-kounterterms-vacuum}
\end{eqnarray}
Proceeding as in the previous section we see that there are no divergent contributions to $<\Delta AF_t^{n-1}\delta _{\lambda }\overline{A} >$ come from 
taking:\\
i. $\Delta A$ along $(P_i-J_{1i})$,
all the $F_t$'s along $J_{ij}$ and $\delta _{\lambda }\overline{A}$ along $(P_i+J_{1i})$ or\\
ii. $\Delta A$ along $(P_i-J_{1i})$, one of the $F_t$'s along 
$(P_i+J_{1i})$, the remaining $F_t$'s but one along $J_{ij}$ and $\delta _{\lambda }\overline{A}$ along $J_{ij}$ or\\
iii. $\Delta A$ along $(P_i+J_{1i})$,
all the $F_t$'s along $J_{ij}$ and $\delta _{\lambda }\overline{A}$ along $(P_i-J_{1i})$ .

Using the explicit forms of the relevant fields and the definition of the symmetrized trace we get 
\begin{eqnarray}
\Omega ^{1}_{2n} (A,\overline{A},\lambda )= \kappa  ~\epsilon _{k_1l_1...k_{n}l_{n}}
\hat{R}^{k_1l_1}_{\infty} ... 
\hat{R}^{k_{n}l_{n}}_{\infty}  \lambda ^{1}+\nonumber\\
+2 n \kappa  \int_0^1dt~(t-1) ~\epsilon _{ijk_1l_1...k_{n-1}l_{n-1}}
\hat{e}_{\infty} ^i \left[ \hat{D}_{\infty}\hat{\lambda} ^{(+)j}\right]
F^{k_1l_1}_{t} ...F^{k_{n-1}l_{n-1}}_{t} +\nonumber\\
+2 n \kappa  \int_0^1dt~(t-1)t ~\epsilon _{ijk_1l_1...k_{n-1}l_{n-1}}
\hat{e}_{\infty} ^i \left[ \hat{D}_{\infty}\hat{\zeta } ^{j}\right]
F^{k_1l_1}_{t} ...F^{k_{n-2}l_{n-2}}_{t} \left[ \hat{D}_{\infty}
\lambda ^{k_{n-1}l_{n-1}} \right]-\nonumber\\
- n \kappa  \int_0^1dt~(t-1) ~\epsilon _{ijk_1l_1...k_{n-1}l_{n-1}}
\hat{\zeta}_{\infty} ^i \left[ \hat{D}_{\infty}\hat{\lambda} ^{(-)j}\right]
F^{k_1l_1}_{t} ...F^{k_{n-1}l_{n-1}}_{t}\label{AdS-gauge-anomaly-kounterterms-1}
\end{eqnarray}
where we ignored contributions along $dr$, as they have no support at the boundary, $F_t^{kl}= \hat{R}^{kl}_{\infty}-t^2(  
\hat{\zeta }^{k}\hat{e}^{l}_{\infty}+\hat{e}^{k}_{\infty}\hat{\zeta }^{l})$, and every field that appears is evaluated at $r\rightarrow\infty$. 

It is possible, integrating by parts, to write the expression for the anomaly in such 
a way that none of the gauge parameters are acted upon by derivatives. As it was said above, given a 
symmetric trace with all indices saturated $<...>$ and a covariant derivative $D$ it holds that 
$d<(something)>=<D(something)>$, where $d$ stands for the exterior derivative. This is true in particular 
for de symmetric trace provided by contraction with the Levi-Civita $\epsilon$-tensor and the covariant derivative
$\hat{D}_{\infty}$. Furthermore $\hat{D}_{\infty}\hat{e}_{\infty} ^i=0$ because of the required vanishing of the intrinsic torsion of the boundary, and $\hat{D}_{\infty}\hat{R}^{ij}_{\infty}=0$ in virtue of the Bianchi identities.
We have
\begin{eqnarray}
\hat{D}_{\infty} F_t^{kl}= -t^2\left[ 
(\hat{D}_{\infty}\hat{\zeta }^{k})\hat{e}^{l}_{\infty}-\hat{e}^{k}_{\infty}(\hat{D}_{\infty}\hat{\zeta }^{l})\right]\\
\hat{D}_{\infty}(\hat{D}_{\infty}\hat{\zeta }^{i})=\hat{R}^{i}_{\infty j}\hat{\zeta }^{j}
\end{eqnarray}
and 
\begin{eqnarray}
d \left[ \epsilon _{ijk_1l_1...k_{n-1}l_{n-1}}
\hat{e}_{\infty} ^i  \hat{\lambda} ^{(+)j}
F^{k_1l_1}_{t} ...F^{k_{n-1}l_{n-1}}_{t} \right] =\nonumber\\
=-\epsilon _{ijk_1l_1...k_{n-1}l_{n-1}}
\hat{e}_{\infty} ^i \left[ \hat{D}_{\infty}\hat{\lambda} ^{(+)j}\right]
F^{k_1l_1}_{t} ...F^{k_{n-1}l_{n-1}}_{t}+\nonumber\\
-(n-1)\epsilon _{ijk_1l_1...k_{n-1}l_{n-1}}
\hat{e}_{\infty} ^i \hat{\lambda} ^{(+)j}
\left[ \hat{D}_{\infty}F^{k_1l_1}_{t}\right]F^{k_2l_2}_{t} ...F^{k_{n-1}l_{n-1}}_{t}
\end{eqnarray}
and also
\begin{eqnarray}
d\{ \epsilon _{ijk_1l_1...k_{n-1}l_{n-1}}
\hat{e}_{\infty} ^i \left[ \hat{D}_{\infty}\hat{\zeta } ^{j}\right]
F^{k_1l_1}_{t} ...F^{k_{n-2}l_{n-2}}_{t} 
\lambda ^{k_{n-1}l_{n-1}} \}=\nonumber\\
-\epsilon _{ijk_1l_1...k_{n-1}l_{n-1}}
\hat{e}_{\infty} ^i \left[ \hat{D}_{\infty}(\hat{D}_{\infty}\hat{\zeta } ^{j})\right]
F^{k_1l_1}_{t} ...F^{k_{n-2}l_{n-2}}_{t} 
\lambda ^{k_{n-1}l_{n-1}} -\nonumber\\
-(n-2)\epsilon _{ijk_1l_1...k_{n-1}l_{n-1}}
\hat{e}_{\infty} ^i \left[ \hat{D}_{\infty}\hat{\zeta } ^{j}\right]
\left[ \hat{D}_{\infty} F^{k_1l_1}_{t} \right]F^{k_2l_2}_{t} ...F^{k_{n-2}l_{n-2}}_{t} 
\lambda ^{k_{n-1}l_{n-1}}  -\nonumber\\
-\epsilon _{ijk_1l_1...k_{n-1}l_{n-1}}
\hat{e}_{\infty} ^i \left[ \hat{D}_{\infty}\hat{\zeta } ^{j}\right]
F^{k_1l_1}_{t} ...F^{k_{n-2}l_{n-2}}_{t} \left[ \hat{D}_{\infty}
\lambda ^{k_{n-1}l_{n-1}} \right] 
\end{eqnarray}
and finally
\begin{eqnarray}
d \left[ \epsilon _{ijk_1l_1...k_{n-1}l_{n-1}}
\hat{\zeta}_{\infty} ^i  \hat{\lambda} ^{(-)j}
F^{k_1l_1}_{t} ...F^{k_{n-1}l_{n-1}}_{t} \right]   =\nonumber\\
=\epsilon _{ijk_1l_1...k_{n-1}l_{n-1}}
\left[ \hat{D}_{\infty}\hat{\zeta}_{\infty} ^i\right] \hat{\lambda} ^{(-)j}
F^{k_1l_1}_{t} ...F^{k_{n-1}l_{n-1}}_{t}-\nonumber\\
-\epsilon _{ijk_1l_1...k_{n-1}l_{n-1}}
\hat{\zeta}_{\infty} ^i \left[ \hat{D}_{\infty}\hat{\lambda} ^{(-)j}\right]
F^{k_1l_1}_{t} ...F^{k_{n-1}l_{n-1}}_{t}-\nonumber\\
-(n-1)\epsilon _{ijk_1l_1...k_{n-1}l_{n-1}}
\hat{\zeta}_{\infty} ^i \hat{\lambda} ^{(-)j}
\left[ \hat{D}_{\infty}F^{k_1l_1}_{t}\right] F^{k_2l_2}_{t}...F^{k_{n-1}l_{n-1}}_{t}
\end{eqnarray}

It is straightforward replace the previous expressions in eq.(\ref{AdS-gauge-anomaly-kounterterms-1}), 
discarding irrelevant total derivatives, to obtain an alternative expression of the AdS gauge anomaly
that has no derivatives of the gauge parameter.
An immediately apparent feature of that alternative expression of the anomaly is that, if the additional condition 
$\hat{D}_{\infty}\hat{\zeta } ^{i}=0$ holds, then only the first term of the second member 
(the first line) is non zero. This condition, which for instance is automatically satisfied if the boundary manifold is of constant curvature, already appeared in ref.\cite{mora-action-principle-cs-ads} as necessary for the finiteness of the action, which otherwise will have a divergence that would be linear on $r$ (or logarithmic in the Fefferman-Graham standard radial coordinate $\rho$).

\subsection{Weyl anomaly}

As said in the previous section, gauge transformations for which only $\lambda ^1$ 
is non vanishing induce Weyl transformations at the boundary. The Weyl anomaly for
the Kounterterms action principle that follows from 
eq.(\ref{AdS-gauge-anomaly-kounterterms-1}) is
\begin{eqnarray}
\Omega ^{1}_{2n} (A,\overline{A},\lambda )= \kappa  ~\epsilon _{k_1l_1...k_{n}l_{n}}
\hat{R}^{k_1l_1}_{\infty} ... 
\hat{R}^{k_{n}l_{n}}_{\infty}  \lambda ^{1}  
= \kappa  E_n  \lambda ^{1} \label{Weyl-anomaly-kounterterms}
\end{eqnarray}
where $E_n =\epsilon _{k_1l_1...k_{n}l_{n}}
\hat{R}^{k_1l_1}_{\infty} ... 
\hat{R}^{k_{n}l_{n}}_{\infty}$ is the Euler density of the boundary 
(which is not required to vanish by the field equations). 
This result agrees with refs.\cite{banados-schwimmer-theisen,banados-olea-theisen,banados-miskovic-theisen}

\subsection{Lorentz and translational anomalies}

From eq.(\ref{AdS-gauge-anomaly-kounterterms-1}) we see that there is no Lorentz anomaly in the Kountereterms approach (unlike what happens in the Backgrounds approach).

The anomalies under gauge translations can be read from eq.(\ref{AdS-gauge-anomaly-kounterterms-1}), but as said above they vanish if 
the condition $\hat{D}_{\infty}\hat{\zeta } ^{i}=0$ holds. This condition was 
found in ref.\cite{mora-action-principle-cs-ads}   to be necessary to ensure the finiteness of the action.

\section{Diffeomorphism Anomalies of  Chern-Simons AdS gravity: Backgrounds}

We will use eqs.(\ref{anomalous-variation-transgression-diffeos-9}) and (\ref{diffeo-anomaly-backgrounds}).
For the AdS vacuum $I_{\xi}<\overline{F}^{n+1}>=0$ and $<\overline{F}^n  I_{\xi}\overline{A}>=0$ as a
result of the field equations. Also the asymptotic dependence of the relevant fields in the backgrounds approach
and the definition of the symmetric trace imply that $<\Delta AF_t^{n-1}I_{\xi}\overline{F}>$. What remains has exactly the same form that the gauge anomaly for backgrounds, but with the 
"effective gauge parameter" $I_{\xi}\overline{A}$ instead of $\lambda$. From
\begin{eqnarray}
I _{\xi } \overline{A}=e^r I _{\xi }  \hat{e}^i_{\infty}(P_i-J_{1i})
+\frac{1}{2}I _{\xi }\hat{\omega}_{\infty }^{ij} J_{ij}+  I _{\xi }( dr)~P_1-
\frac{1}{2}e ^{-r}I _{\xi }\hat{\overline{\zeta}}^i(P_i+J_{1i})
\end{eqnarray}
we can read the corresponding components of that "effective $\lambda$" and just replace them
in eq.(\ref{AdS-anomaly-backgrounds}) to obtain the diffeomorphism anomaly for 
Chern-Simons AdS gravity in the backgrounds approach. One important difference is that the components of this effective gauge parameter are not all independent, as once $\hat{e}^i_{\infty}$ is given then $\hat{\omega}_{\infty }^{ij}$ and
then $\hat{\overline{\zeta}}^i$ are determined.

\section{Diffeomorphism Anomalies of  Chern-Simons AdS gravity: Kounterterms}

Our starting point are eqs.(\ref{anomalous-variation-transgression-diffeos-9}) and (\ref{diffeo-anomaly-backgrounds}). The first observation is that for the Kounterterms vacuum $I_{\xi}<\overline{F}^{n+1}>=0$, as $\overline{F}$ only has components along $J_{ij}$.
We have that $I_{\xi}\overline{F}=\frac{1}{2}I_{\xi}\hat{R}_{\infty}^{ij}J_{ij}$ and
$I_{\xi}\overline{A}=\frac{1}{2}I_{\xi}\hat{\omega}_{\infty}^{ij}J_{ij}+ I_{\xi}(dr)P_1$.
Looking at each term of $\Omega ^1_{2n}(A,\overline{A},\xi)$ we see that:\\
i. $<\overline{F}^n  I_{\xi}\overline{A}>$ is non zero if the
$\overline{F}$'s are along $J_{ij}$ and $I_{\xi}\overline{A}$ along $P_1$.\\
ii. $<\Delta AF_t^{n-1}I_{\xi}\overline{F}>$ has a non zero contribution if 
$\Delta A$ is along $(P_i-J_{1i})$, one $F_t$ is along $(P_i+J_{1i})$, 
the remaining 
$F_t$'s along $J_{ij}$, and $I_{\xi}\overline{F}$ along $J_{ij}$.\\
iii. $<\Delta AF_t^{n-1}\overline{D}I_{\xi}\overline{A}>$ appears in a term that 
has exactly the same form that the last term in the gauge anomaly, with $I_{\xi}\overline{A}$ instead of $\lambda$. It follows that the discussion done for gauge anomalies apply. Notice that the "effective gauge parameter" $I_{\xi}\overline{A}$ does not contain gauge translations, and that the non vanishing components have the same asymptotic dependence that we required for $\lambda $.

Putting all together, the explicit form of the diffeomorphism anomaly is
\begin{eqnarray}
\Omega ^{1}_{2n} (A,\overline{A},\xi )= \kappa  ~\epsilon _{k_1l_1...k_{n}l_{n}}
\hat{R}^{k_1l_1}_{\infty} ... 
\hat{R}^{k_{n}l_{n}}_{\infty} I_{\xi}(dr) 
+\nonumber\\
+2 n \kappa  \int_0^1dt~(t-1)t ~\epsilon _{ijk_1l_1...k_{n-1}l_{n-1}}
\hat{e}_{\infty} ^i \left[ \hat{D}_{\infty}\hat{\zeta } ^{j}\right]
F^{k_1l_1}_{t} ...F^{k_{n-2}l_{n-2}}_{t} \left[ I_{\xi}\hat{R}_{\infty}^{k_{n-1}l_{n-1}} \right]+\nonumber\\
+2 n \kappa  \int_0^1dt~(t-1)t ~\epsilon _{ijk_1l_1...k_{n-1}l_{n-1}}
\hat{e}_{\infty} ^i \left[ \hat{D}_{\infty}\hat{\zeta } ^{j}\right]
F^{k_1l_1}_{t} ...F^{k_{n-2}l_{n-2}}_{t}  \hat{D}_{\infty}(I_{\xi}
\hat{\omega}_{\infty} )^{k_{n-1}l_{n-1}} 
\label{diffeo-anomaly-CS_AdS-kounterterms}
\end{eqnarray}
Notice that, as it was the case for gauge anomalies, 
if the condition $\hat{D}_{\infty}\hat{\zeta }_{\infty} ^{i}=0$ holds the 
only term of the second member that survives is the first. That term corresponds
to the Weyl anomaly, as $I_{\xi}(dr)$ is non vanishing only for 
$\xi =\sigma \frac{\partial ~}{\partial r}$ where $\sigma$ is an infinitesimal 
parameter (or function) for infinitesimal coordinate transformations, and it is 
well known \cite{imbimbo}, and easy to check, that radial diffeomorphisms induce Weyl 
transformations at the boundary. In that case
$I_{\xi}(dr)=\sigma $ and the Weyl anomaly is 
\begin{eqnarray}
\Omega ^{1}_{2n} (A,\overline{A},\xi )= \kappa  ~\epsilon _{k_1l_1...k_{n}l_{n}}
\hat{R}^{k_1l_1}_{\infty} ... 
\hat{R}^{k_{n}l_{n}}_{\infty} \sigma 
\label{weyl-diffeo-anomaly-CS_AdS-kounterterms}
\end{eqnarray}
again in agreement with refs.\cite{banados-schwimmer-theisen,banados-olea-theisen,banados-miskovic-theisen}.

\section{Discussion and Comments}

We have computed AdS gauge and diffeomorphism anomalies for Chern-Simons AdS gravity, for two action principles
discussed in previous work \cite{motz1,motz3,mora-action-principle-cs-ads}, and assuming the asymptotic
behavior of the fundamental fields proposed in \cite{mora-action-principle-cs-ads}.
The anomalies arise as a result of regarding the second field configuration, regarded a regulator,
as non varying. These means in particular that if one regards both $A$ and $\overline{A}$ as dynamical
fields varying with the same transformation rules there would be no anomalies at all.

The results are finite without requiring further regularization or subtraction, adding to the evidence that
the action principles motivated by the transgression are indeed the appropriate ones. 

Anomalies are characterized by a symmetric tensor (or equivalently a symmetrized trace) in the algebra of the 
relevant gauge group as an overall constant factor (see for instance \cite{wu-zee-zumino,alvarez}). The anomalies 
found here, in particular Lorentz otr gravitacional anomalies, involve the Levi-Civita tensor as said invariant trace. 
On the other hand standard gravitational anomalies for theories with chiral fermions with 
the standard minimal coupling to gravity are given in terms of standard traces 
properly symmetrized \cite{bardeen-zumino,alvarez}.
This is somewhat puzzling, as it is not clear what sort of 
hypothetical dual CFT theory would yield matching anomalies. 

Gravitational anomalies of the standard form have been studied in the AdS-CFT 
correspondence context, for instance in \cite{kraus-larsen,solodukhin}. However these anomalies
are generated by adding Lorentz Chern-Simons terms in the bulk, constructed with a symmetrized standard trace
instead of the Levi-Civita tensor. It seems that to generate gravitational anomalies
of the standard form one should start with a lagragian of the kind known as "exotic Chern-Simons gravity", where
the invariant tensor in the AdS gauge group algebra is a symmetrized combination of standard traces, instead of 
the Levi-Civita tensor. It would be interesting to do an analysis similar to \cite{mora-action-principle-cs-ads}
for "exotic CS gravities" and then compute the AdS anomalies that could arise in that case.

Another interesting set of questions has to do with the role of torsion in the anomalies computed. The analysis
of \cite{mora-action-principle-cs-ads} showed that the intrinsic torsion of the boundary must vanish if the 
AdS gauge curvature is asymptotically finite. However the bulk torsion itself is not required to vanish, 
and for instance $\zeta ^i$ contains information about it. It would be valuable 
to understand better the physical implications of the bulk torsion contribution to the anomalies computed.
It would be interesting to see if the anomalies discussed here are related to the ones discussed in \cite{blagojevic},
for a different kind of theory and in 2+1 dimensions. 

Anomalies involving space-time torsion of the kind introduced 
in \cite{chandia-zanelli-1,chandia-zanelli-2} are excluded in our framework, as we
have a vanishing boundary torsion as an asymptotic condition and furthermore and 
those anomalies involve standard traces.
However they may appear in a holographic context involving exotic CS gravities, 
which in addition to use standard traces 
may admit a different asymptotic torsion.

The possible forms of anomalies are quite constrained by the Wess-Zumino (WZ) consistency 
conditions (see for instance \cite{wu-zee-zumino,bardeen-zumino}), what means that anomalies
of the form discussed here may appear may be relevant for other field theories.

In particular, they may be relevant for Lovelock gravities\footnote{Lovelock gravities \cite{lovelock-1,lovelock-2} are a class of gravity theories  that were shown \cite{zwiebach,zumino} to be possible low energy effective descriptions of gravity in string theory. These theories have been extensively studied afterwards.} and their holographic duals. 
It has been proved that the boundary term suggested by transgressions \cite{motz1,motz3}, with a suitable constant coefficient, also works as a regulator for General Relativity with a cosmological constant in odd dimensions \cite{motz2}, and in 
fact for any Lovelock AdS gravity \cite{olea-kounterterms,kofinas-olea}. That is surprising at first, 
but maybe the fact that the variation of that boundary term (which in fact has information about the bulk) 
would generate anomalies satisfying the WZ consistency condition may explain that fact.

{\bf Acknowledgments:} I am grateful to O. Miskovic and R. Olea for enlightening discussions and comments.
I acknowledge financial support from the {\it Sistema Nacional de Investigadores} (SNI), Uruguay, while part of this work
was done.

\appendix

\end{document}